\documentclass{aa}
\usepackage{graphicx}
\usepackage{txfonts}
\usepackage{natbib}
\begin{document}

   \title{Sun-as-a-star observations: evidence for degree dependence of changes in
   damping of low-$\ell$ p modes along the solar cycle}

    \author{D.~Salabert\inst{1,2}, W.~J.~Chaplin\inst{3}, Y.~Elsworth\inst{3},
    R.~New\inst{4} and G.~A.~Verner\inst{3}}

    \offprints{D.~Salabert, \email{dsalabert@nso.edu}}

    \titlerunning{Degree dependence of low-$\ell$ p-mode parameters with solar activity}
    \authorrunning{Salabert et al.}

    \institute{National Solar Observatory,
               950 North Cherry Avenue, Tucson, AZ 85719, USA
          \and
               High Altitude Observatory, National Center for Atmospheric Research, P.O. Box 3000,
               Boulder, CO 80307-3000, USA
          \and
               School of Physics and Astronomy,
               University of Birmingham, Edgbaston,
               Birmingham B15 2TT, UK
          \and
               Faculty of Arts, Computing, Engineering and Sciences,
               Sheffield Hallam University, Sheffield S1 1WB, UK
          }

\date{Received xxxxxxx xx, xxxx / Accepted xxxxxxx xx, xxxx}

\abstract
       {}
       {We use 9.5-yr of BiSON Sun-as-a-star data to search for
dependence of solar-cycle parameter changes on the angular degree,
$\ell$, of the data. The nature of the Sun-as-a-star observations is
such that for changes measured at fixed frequency, or for changes
averaged across the same range in frequency, any $\ell$ dependence
present carries information on the latitudinal distribution of the
agent (i.e., the activity) responsible for those changes.}
      {We split
the 9.5-yr timeseries into contiguous 108-d pieces, and determine
mean changes in the damping of, power in, and energy supplied to the
modes through the solar cycle. We also apply a careful correction to
account for the deleterious effects of the ground-based BiSON window
function on the results.}
      {From our full analysis we obtain a
marginally significant result for the damping parameter, where the
mean change is found to be weakest at $\ell=0$. The other parameters
show hints of some dependence in $\ell$.}
      {Our main conclusion is that
the mean fractional solar-cycle change in the $\ell=0$ damping rates
is approximately 50\,\% smaller than was previously assumed. It had
been common practice to use an average over all low-$\ell$ modes;
our downward revision of the radial-mode value has implications for
comparisons with models of the global solar cycle changes, which are
usually based on a spherically symmetric geometry.}


   \maketitle

\section{Introduction}
\label{sec:intro}

The fact that damping rates and powers of the global p modes change
through the solar cycle is now well established \citep[e.g.,
][]{chaplin00,komm00,salabert03,chano03,chano04,salabert06}. At
least where the main part of the p-mode spectrum is concerned
damping gets heavier, and observed power gets weaker, as the level
of solar activity increases. Information on the damping and power
parameters comes straightforwardly from the observations. The
damping rates are assumed to be linearly related to the linewidths
of the resonant peaks in the frequency power spectrum; while the
powers are proportional to the product of the peak widths and
heights (the latter more formally termed the maximum power spectral
densities).

\citet{komm02} took advantage of the large number of components
available to analyze in the medium-degree range (from $\ell=40$ to
80) the latitudinal dependence of the width and height
changes. They found that the changes were concentrated in latitudes
occupied by the active regions. The damping rates and powers, like
the mode frequencies, therefore seemed to be responding in some
fashion to changes wrought on the active regions by the changing
magnetic fields.

Most of the low-$\ell$ results have come from the Sun-as-a-star data
(e.g., from BiSON, GOLF, IRIS and VIRGO/SPM; or from GONG and MDI
data combined into a Sun-as-a-star-like proxy). These data show only
the even $\ell+m$ components; the odd components are so weak as to
be unobservable. Since it is the outer, sectoral components (with
$\ell=|m|$) that appear most prominently, the Sun-as-a-star mode
parameters estimated by the usual analysis methods are dominated by,
and therefore \emph{close to}, the sectoral values. This
characteristic gives a noticeable change in the latitudinal
sensitivity from $\ell=0$ to $\ell=1$, 2 or 3. The spatial
sensitivity of Sun-as-a-star data at the latter three values of
$\ell$ is weighted toward the lower latitudes, where the active
regions reside. The Sun-as-a-star parameters at these $\ell$ are
therefore more sensitive to the solar cycle than are the $\ell=0$
data. Evidence for $\ell$ dependence in the shifts of the
Sun-as-a-star mode parameters therefore carries information on the
latitudinal distribution of the agent responsible for those shifts.
(Changes in inertia, at fixed frequency, between these $\ell$ are
very small indeed.)

It is much more difficult to uncover spatial dependence of the mode
parameter changes in the low- than in the medium-$\ell$ data because
there are far fewer components to analyze, and uncertainties on the
results are commensurately larger. When changes to damping and power
were first uncovered in the Sun-as-a-star data, results were
averaged over $\ell$ to reduce errors. Given the modest precision in
the results, it was assumed the $\ell$-averaged values also provided
a working proxy of the radial-mode shifts. And so these averages
were used as meaningful comparisons for models of global changes in
damping \citep[see, e.g.,][]{houdek01}, models set up in spherical
geometry, i.e., pertinent only to the $\ell=0$ case. With better
analysis, and more data, it has become possible to give results for
each $\ell$, and to therefore test whether the $\ell=0$ shifts
really are weaker than an average across $\ell=0$ to 2 or 3.

We made a first cut at such an analysis in \citet{chaplin03a}. Our
results, from BiSON Sun-as-a-star data, suggested very strongly that
the $\ell=0$ linewidth changes were indeed significantly weaker than
at $\ell=1$ and 2. The clear implication was that results of the
theoretical models would now need to be compared to this new,
smaller shift. In the analysis of the BiSON data, we had to allow
for the corrupting influence on the results of the ground-based
BiSON window function. We did so by applying corrections that were
designed originally for a different study, using data from a
different period. In this paper we revisit our analysis. We design
and implement a correction procedure for the 9.5-yr BiSON dataset in
question. We find that implementation of this internally consistent,
and more accurate, correction gives little change to the results,
and reinforces our earlier conclusion. In summary, the mean
fractional solar-cycle change in the $\ell=0$ damping rates is
approximately 50\,\% smaller than was previously assumed.

 \section{Data analysis}
 \label{sec:dataanal}

 \subsection{Observations}

Our analysis made use of 3456\,d of data on the low-angular-degree
solar p-modes, collected by the six-station, ground-based Birmingham
Solar-Oscillations Network (BiSON). This 9.5-yr dataset had a
starting date of 1992 July 26, and spanned the falling phase of
solar cycle 22 and the rising phase, and the maximum, of solar cycle
23.

The instruments at each BiSON site make Sun-as-a-star observations
of the Doppler shift of the potassium Fraunhofer line at 770\,nm
\citep[e.g., ][]{chaplin96}. Raw data were first processed in the
manner described by \citet{elsworth95} to yield daily calibrated
velocity residuals.  We then combined coherently the resulting
$\approx 2 \times 10^4$ individual daily sets from all six stations
to yield the principal 3456-d time series of residuals. The duty
cycle of this combined set was $\sim 76$\,\%, with breaks in
coverage largely the result of inclement weather.

In order to derive solar-cycle trends, the 3456-d series was divided
in thirty-two contiguous, independent 108-day pieces, which had a
range of duty cycles between 63 and 87 per cent. This 108-d
interval, a multiple of the solar rotation period, was chosen to
minimize effects dependent on the solar rotation.

 \subsection{Mode parameter extraction}
 \label{sec:extrac}

The power spectrum of each 108-day time series was fitted to yield
estimates of the mode parameters over a range in frequency $1600 \le
\nu \le 3800\,\rm \mu Hz$. This fitting was done by a multi-step
iterative method based on \citet{gelly02}. The asymmetric profile of
\citet{nigam98} was used to describe each component, as:
\begin{equation}
  \mathcal{M}(x)=H\frac{(1+b x)^2+b^2}{1+x^2},
 \label{eq:asym}
 \end{equation}
where $x=2(\nu-\nu_0)/\Gamma$, and $H$, $\Gamma$, $\nu_0$ are the
mode height (maximum power spectral density), the \textsc{fwhm}
linewidth, and the central frequency respectively. The parameter $b$
is the asymmetry coefficient, which in the present analysis was
fixed at small negative values determined by analysis of longer
datasets.

Due to their close proximity in frequency, modes were fitted in
pairs (i.e., $\ell=2$ with 0 and $\ell=3$ with 1). An additional
offset was included in the fitting model to describe the background
level in the fitting window. The first temporal sidebands at
$11.57\,\rm \mu Hz$ were also included in the model. The mode
parameters were extracted by maximizing a likelihood function
commensurate with the $\chi^2$, 2 degrees-of-freedom statistics of
the power spectrum. The natural logarithm of the height, width and
background terms were varied -- not the parameter values themselves
-- in order to give quasi-normal fitting distributions. Formal
uncertainties on the fitted values were then derived from the
Hessian matrix of each fit in the usual manner.

  \subsection{Damping and excitation parameters}

A generally accepted analogy for the p modes is a forced and damped
harmonic oscillator. For any given mode ($n,\ell$), the mode width,
$\Gamma_{n,\ell}$, is then linearly related to its damping rate,
$\eta_{n,\ell}$. The total velocity power, $P_{n,\ell}$, gives a
measure of the balance between the excitation and damping, and is
proportional to the product of the height, $H_{n,\ell}$, and the
mode width, $\Gamma_{n,\ell}$. The rate at which the energy is
supplied to the modes, $\dot{E}_{n,\ell}$, is a direct measure of
the net forcing and is independent of the mode damping. This `supply
rate' can be computed as the product of the mode width,
$\Gamma_{n,\ell}$, and the mode energy, $E_{n,\ell}$, the mode
energy being proportional to the mode power, $P_{n,\ell}$. For a
detailed description of the mode damping and excitation parameters,
see, e.g., \citet{chaplin00}.

Since the peak-finding procedure returned natural logarithms of the
mode heights and widths, estimates of the mode velocity power and
supply rate were made by taking the combinations:
\begin{equation}
  \log_e(P_{n,\ell})=\log_e(H_{n,\ell})+\log_e(\Gamma_{n,\ell})+\rm {c_{1}},
\label{eq:logpower}
\end{equation}
and
\begin{equation}
  \log_e(\dot{E}_{n,\ell})=\log_e(H_{n,\ell})+2\log_e(\Gamma_{n,\ell})+\rm c{_{2}}.
\label{eq:logesupply}
\end{equation}
Variations in the fitted logarithmic parameters therefore
corresponded to fractional variations of the absolute parameters.
The factors $\rm {c_{1}}$ and $\rm c{_{2}}$ in the above contain
information on the mode inertia, and mode visibility given by the
instrument. The factors were assumed to remain unchanged over the
epoch in question, and were therefore ignored in the solar-cycle
analysis. Uncertainties on the mode velocity power and on the mode
energy supply rate were determined by taking into account the strong
anti-correlation between the extracted mode heights and mode widths
\citep{chaplin00}.

Hereafter, we write the natural logarithm of $H_{n,\ell}$,
$\Gamma_{n,\ell}$, $P_{n,\ell}$, and $\dot{E}_{n,\ell}$ as
$h_{n,\ell}$, $\gamma_{n,\ell}$, $p_{n,\ell}$, and
$\dot{e}_{n,\ell}$ respectively.

 \subsection{Determination of the fractional variations}
 \label{sec:fitflux}

Estimates of the fractional variations of the mode parameters over
time were obtained by performing a linear regression of each of the
fitted parameters $\gamma_{n,\ell}$, $p_{n,\ell}$, and
$\dot{e}_{n,\ell}$, on the 10.7-cm solar radio flux\footnote{Data
available at {\tt http://spidr.ngdc.noaa.gov/spidr/}.} ($F_{10.7}$).
The radio flux was used as a robust proxy of the global level of
solar activity. The best-fitting gradients therefore gave for each
mode a measure of sensitivity of its parameters to the solar
activity.

The regressions were computed as both un-weighted and weighted
linear fits. The formal parameter uncertainties from the
mode-fitting procedures, derived from the Hessian matrices of the
fits, were used here to compute weights for weighted linear
regressions. Fitted gradients from the regressions were then
averaged at each degree over the frequency range between
2600~$\mu$Hz and 3600~$\mu$Hz. This averaging yielded estimates of
\emph{mean} mode changes along the cycle.

There are different ways to compute these averages and their
associated uncertainties. On the one hand, a simple,
\emph{unweighted mean} of the input residuals may be calculated,
with the scatter on the residuals used to determine the uncertainty
on the mean (in the usual manner). An alternative is to use the
Hessian-determined uncertainties on the best-fitting gradients to
calculate weighted means. The error on the mean can then be
estimated in one of two ways: first, from a suitable combination of
the Hessian-determined uncertainties, giving an \emph{internal
error}; and second, by using the Hessian-determined uncertainties to
weight a calculation of the dispersion of the data, giving an
\emph{external error} \citep{chaplin98}.

Since the regression fits were performed in two ways (un-weighted
and weighted) there are six ways in all to estimate the degree
dependence of the p-mode damping and excitation parameters, and the
associated uncertainties (although the weighted-average methods of
course give the same means, but have different errors).

 \section{Correction for bias}
 \label{sec:bias}

Gaps present in the timeseries of a ground-based window function
lead to overestimation of the mode widths and underestimation of the
mode heights \citep{chaplin03b}. Proper allowance must therefore
be made for any bias introduced by the window functions of the
thirty-two 108-d BiSON timeseries. Furthermore, marked changes in
the intrinsic quality of the data can in principle influence the
parameters, via the impact changes in quality may have on the
observed height-to-background ratios of the mode peaks. Changes in
height-to-background ratio must, however, be large to give any
significant bias in the parameters.

\begin{figure}[t]
\centering
\resizebox{\hsize}{!}{\includegraphics{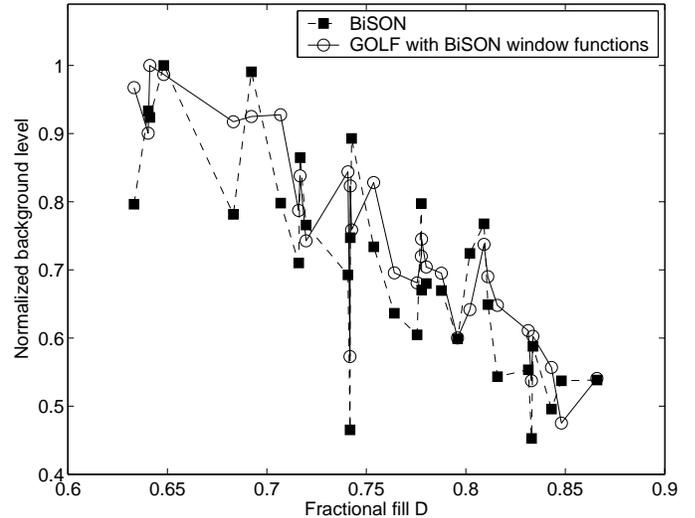}}
\caption{Normalized background level for the thirty-two 108-d BiSON
spectra (-$\blacksquare$-) and the single 108-d GOLF set {\it
modulated} in turn by each of the BiSON window functions
(-$\circ$-). Background levels are averaged over 2600 and
3600~$\mu$Hz, and normalized so that the highest datum takes the
value unity. Data are plotted as a function of the duty cycle, $D$.}
\label{fig:normnoise}
\end{figure}

We have used Monte Carlo simulations of artificial Sun-as-a-star
data to help us correct the BiSON results for bias introduced by
these effects. We found that the bias was dominated by the impact of
the different 108-d window functions; the impact of changes in data
quality was far less significant. The domination of the window
function is illustrated in Fig.~\ref{fig:normnoise}. It shows the
mean background levels, of the thirty-two 108-d BiSON, spectra over
the range 2600 to $3600\,\rm \mu Hz$ (-$\blacksquare$-). The values
have been scaled so that the highest mean background level is set to
unity. The levels are plotted as a function of the duty cycle, $D$,
of each of the sets. As fill decreases, the redistribution of power
by the window function increases the background level.

Also shown in Fig.~\ref{fig:normnoise} are results for a single,
108-d GOLF timeseries. The GOLF timeseries has nearly a 100\,\% duty
cycle, and has been modulated, in turn, by each of the thirty-two
BiSON window functions. The resulting mean background levels have
been plotted (-$\circ$-). Because the GOLF data are the same for
each computation, the change in the background level comes wholly
from the impact of the window function. Changes to the background in
these modulated GOLF data are seen to follow very closely the
changes observed in the BiSON data. From this, we conclude the
window function dominates changes to the background. The result
simplifies the generation of artificial data for the Monte Carlo
simulations in that we do not need to worry about introducing
changes in intrinsic data quality across the simulated datasets.


 \subsection{Bias estimation from artificial Sun-as-a-star data}
 \label{sec:art}

The artificial data were generated using the solarFLAG\footnote{FLAG
URL: http://bison.ph.bham.ac.uk/\,\~\,wjc/Research/FLAG.html}
\citep{chaplin06} mode simulation code. The Laplace transform
solution of the equation of a forced, damped harmonic oscillator was
used to generate each component at a 40-s cadence in the time
domain, in the manner described by \citet{chaplin97}. Components
were re-excited independently at each sample with small `kicks'
drawn from a Gaussian distribution. The model gave rise to peaks in
the frequency power spectrum whose underlying shapes were
Lorentzian. It should be pointed out that the observed low-$\ell$
peaks are slightly asymmetric in shape (albeit at the level of only
a few per cent at most).

Datasets were constructed component by component in the time domain.
Sets were made with a full cohort of simulated low-$\ell$ modes,
covering the ranges $0 \le \ell \le 5$ and $1000 \le \nu \le
5500\,\rm \mu Hz$. A database of mode frequency, power and linewidth
estimates, obtained from analyses of GOLF and BiSON data, was used
to guide the choice of input values for time-series construction. A
rotational frequency splitting of $0.4\, \rm \mu Hz$ was imposed on
all the non-radial modes, irrespective of overtone number and
angular degree, to match that extracted from observations. A `pink'
background noise component was also added in the time domain, the
size of which increased at lower frequencies in order to give
signal-to-background ratios commensurate with GOLF or BiSON-like
Doppler velocity data.

Two cohorts of data were made, both comprising twenty 3456-d
timeseries. Each timeseries was excited with independent realization
noise. We introduced into the first cohort (\#\,1) systematic
changes in the frequency, damping and power parameters so as to
simulate the effects of the solar cycle \citep[see][]{chaplin06}.
These changes were based on variations of a modelled proxy of the
10.7-cm radio flux. The proxy was made to mimic the radio flux for
the real epoch covered by the BiSON timeseries. Parameter variations
were calibrated to give equal, and oppositely directed, changes in
the damping and power parameters (i.e., $\delta \gamma_{n,\ell} =
\delta p_{n,\ell}$), to match previous observations 
\citep[e.g.,][]{chaplin00}. In this scenario 
the supply rate, $\dot{e}_{n,\ell}$,
shows no net change. The simulated changes were frequency dependent,
showing their largest variations at 2900~$\mu$Hz. Furthermore, they
were given \emph{no} $\ell$ dependence. In contrast, in the second
`stationary' cohort (\#\,2) all mode parameters were held constant
in simulated time.

We analyzed each artificial 3456-d dataset in the manner described
in Sections~\ref{sec:extrac} and~\ref{sec:fitflux}, the first step
being to split them into 108-d pieces. From the analysis, we
obtained $\ell$-dependent estimates of mean changes in the
$\gamma_{n,\ell}$, $p_{n,\ell}$, and $\dot{e}_{n,\ell}$ parameters.
Results from all 3456-d sets were then averaged within their
respective cohort to reduce errors. Final estimates of the mean
parameter changes could then be compared with the true changes
introduced, to allow for a determination of the bias.

\begin{figure*}
\centering
\resizebox{\hsize}{!}{\includegraphics{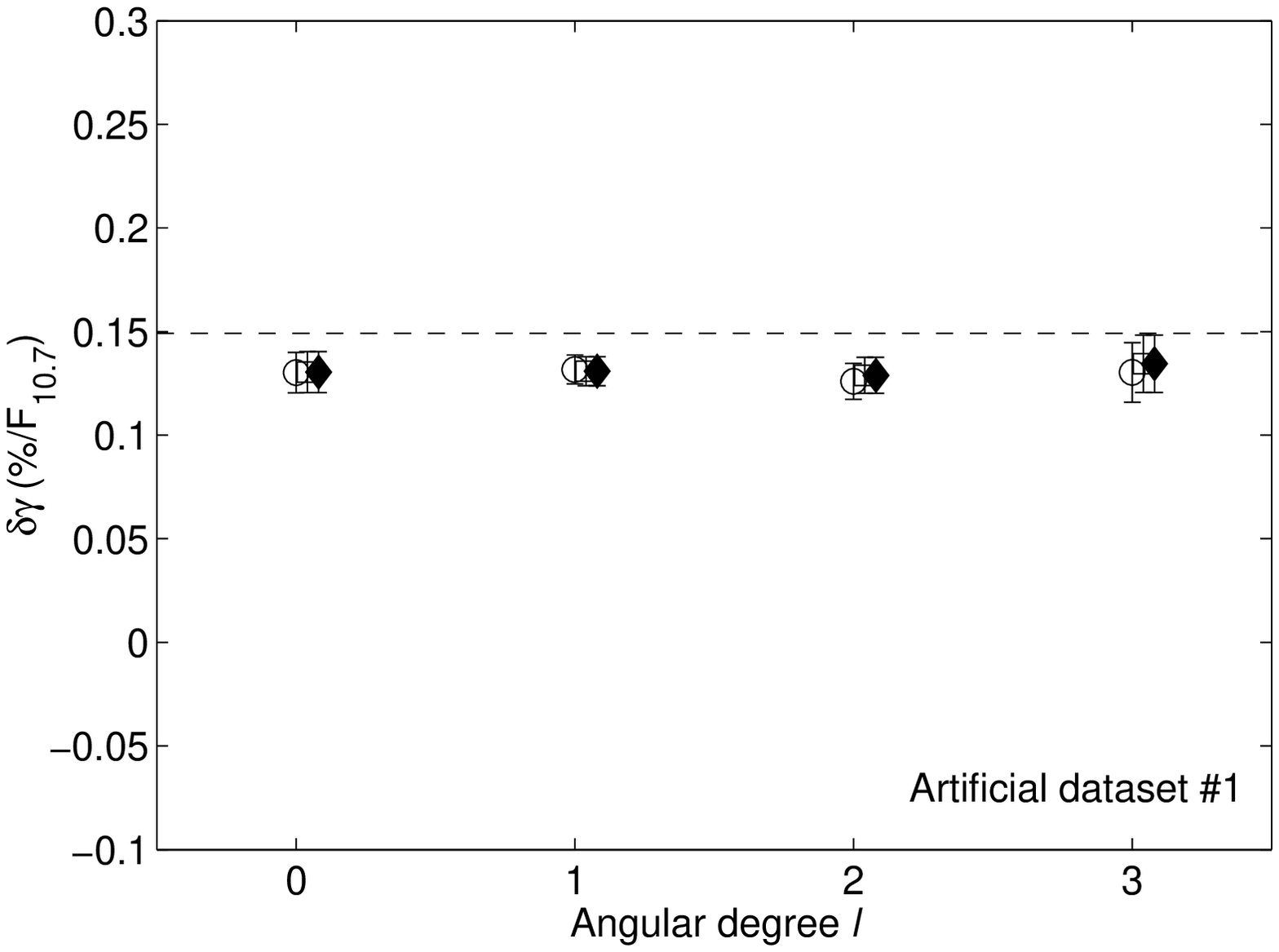}
                      \includegraphics{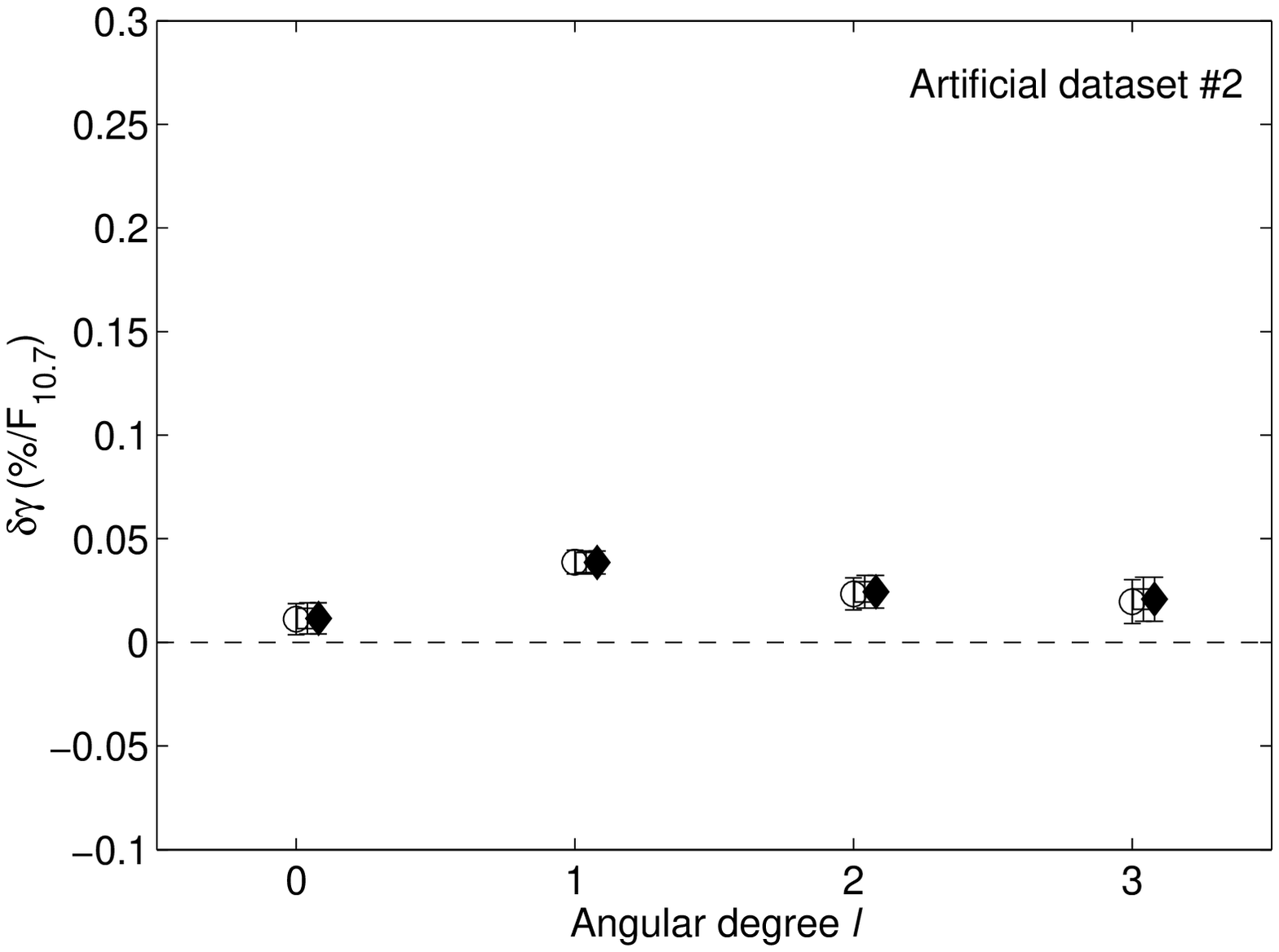}}
\resizebox{\hsize}{!}{\includegraphics{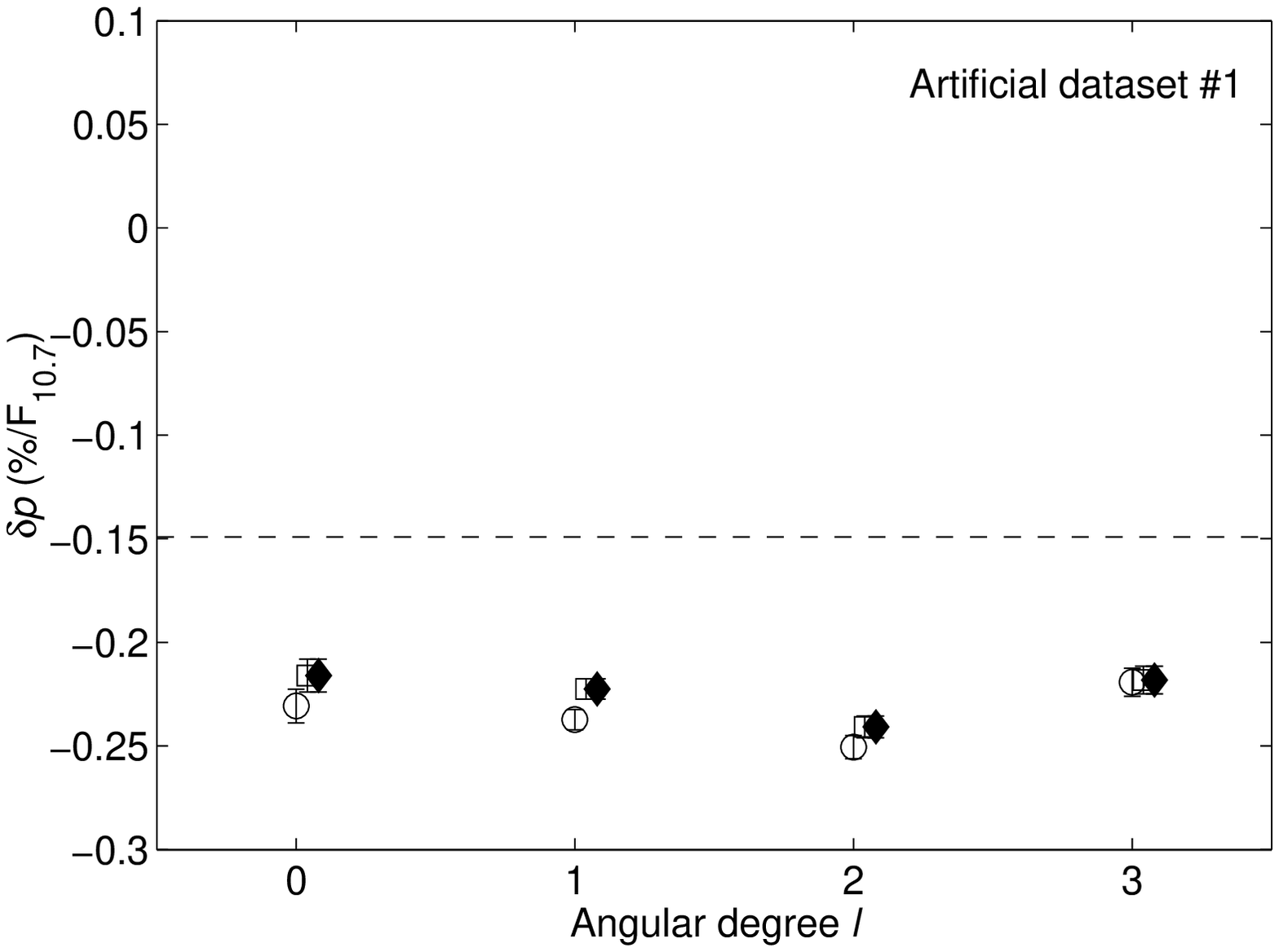}
                      \includegraphics{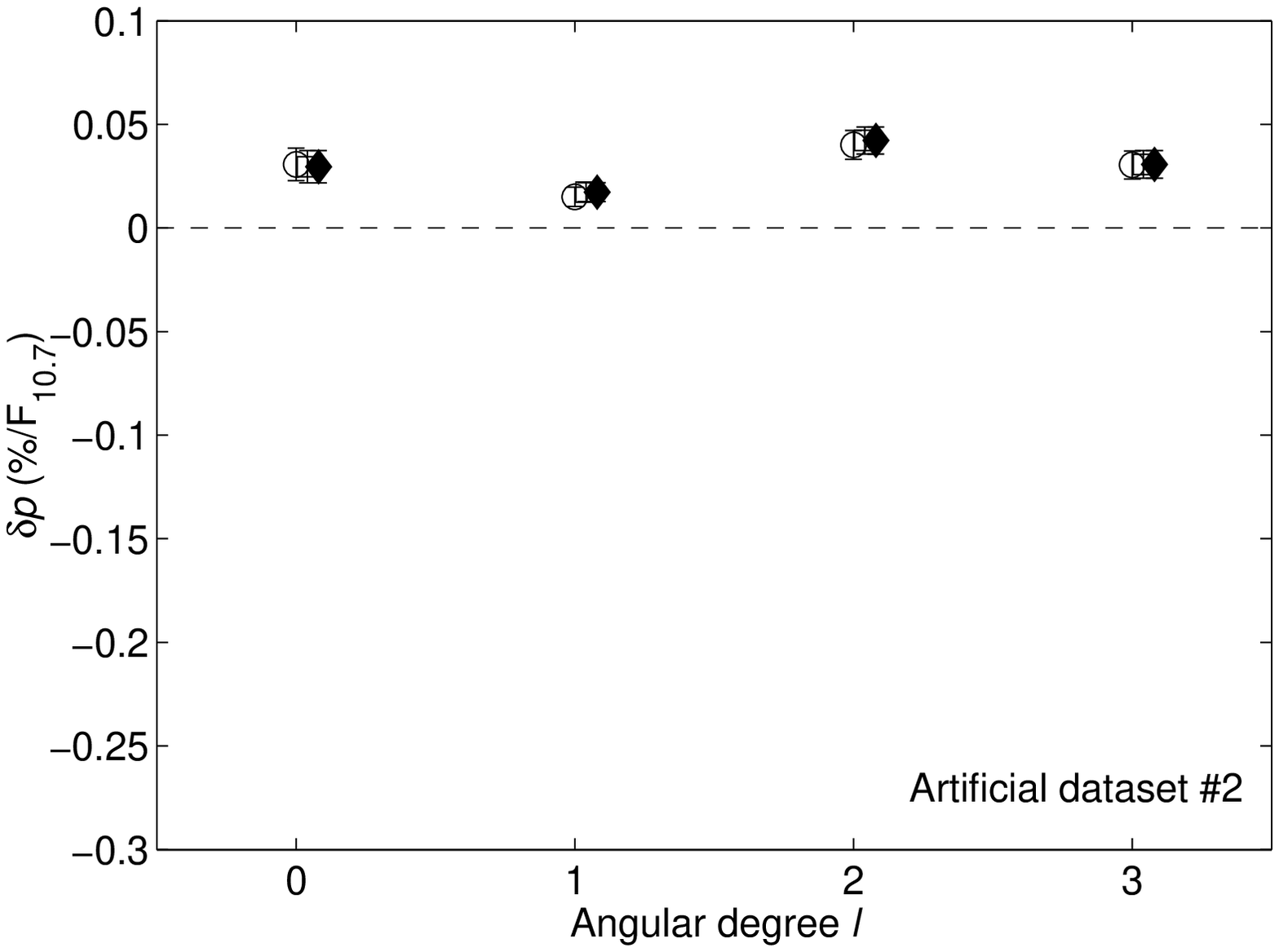}}
\resizebox{\hsize}{!}{\includegraphics{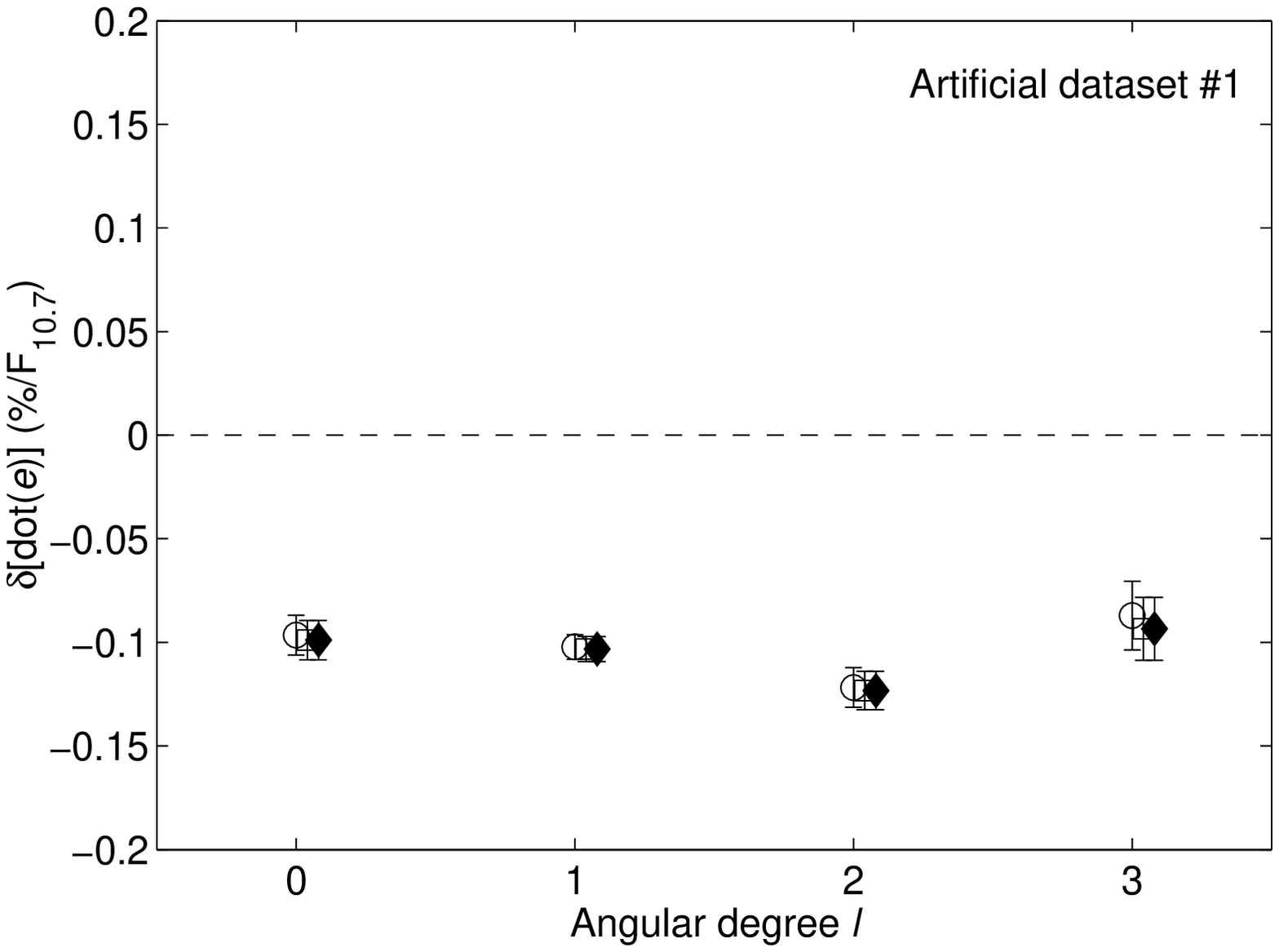}
                      \includegraphics{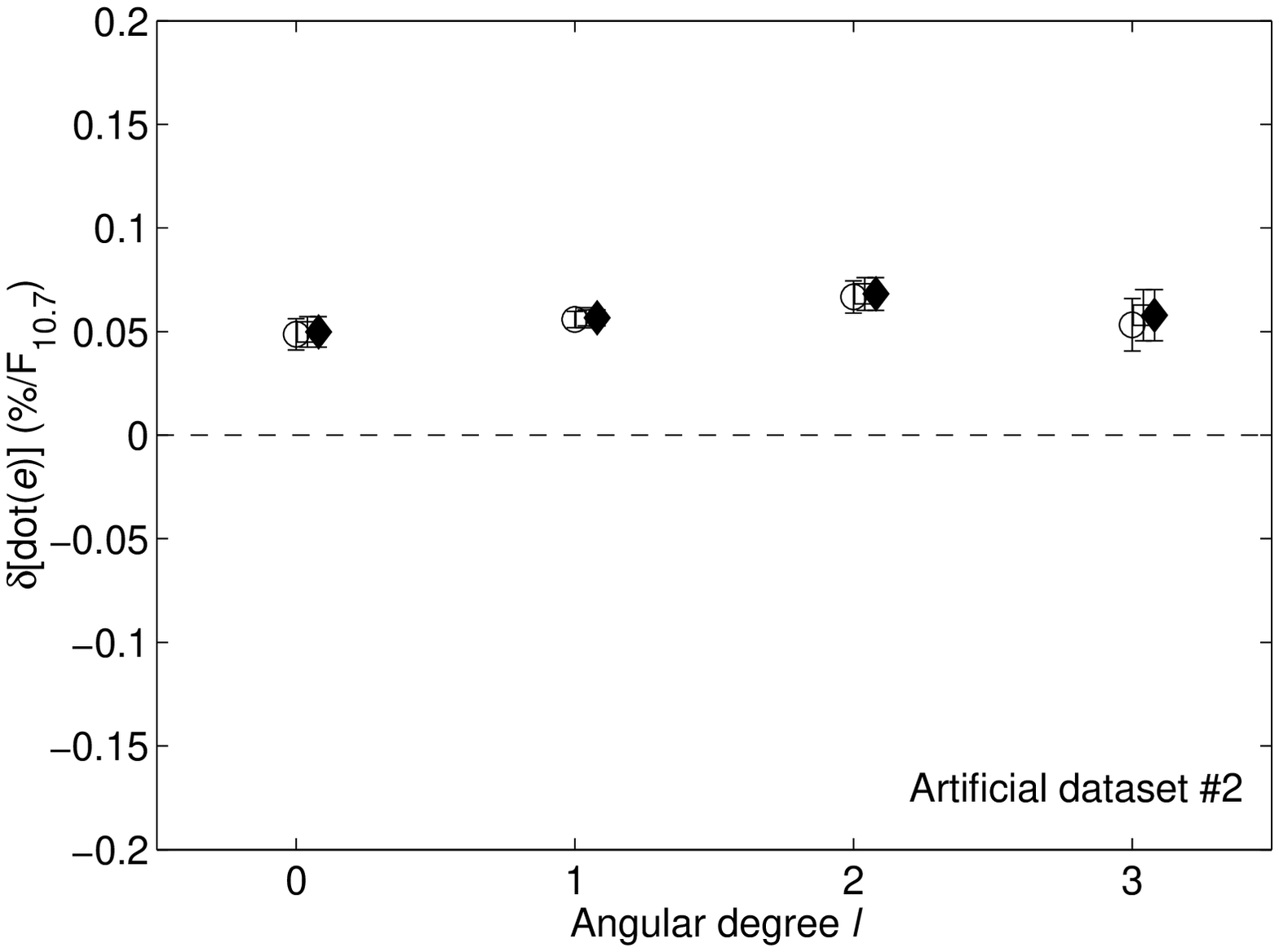}}

\caption{Results of analysis of artificial data. Shown are mean
fractional variations (\% per unit of change in $F_{10.7}$) in mode
damping, $\delta \gamma$ ({\it upper panels}), mode velocity power,
$\delta p$ ({\it middle panels}), and mode energy supply rate,
$\delta \dot{e}$ ({\it lower panels}), as a function of the mode
angular degree $\ell$. Panels in the left-hand column show results
averaged over the twenty artificial 3456-d sets in cohort\,\#\,1;
those in the right-hand column similar averages over sets in
cohort\,\#\,2. The dashed lines in each plot represent the modelled,
input changes. The symbols give information on how the averages were
computed (see Sec.~\ref{sec:fitflux}): unweighted mean -$\circ$-;
weighted mean with internal error -$\square$-; weighted mean with
external error -$\blacklozenge$-. Plotted errors are those on the
mean of the twenty artificial datasets.}

\label{fig:artinocorr}
\end{figure*}

Similar mean values and errors were returned by the different
averaging methods (Section~\ref{sec:fitflux}). For the sake of
clarity we show in what follows results from the weighted
regressions only. The three panels in the left-hand column of
Fig.~\ref{fig:artinocorr} show the mean estimated parameter
variations, per unit change in the radio flux, given by analysis of
artificial cohort\,\#\,1 (the `solar-cycle' data). The right-hand
panels show results from cohort\,\#\,2 (the `stationary' data). The
horizontal dashed lines show the actual mean changes that were
introduced in the artificial parameters. The plotted uncertainties
in each panel show the error bars on the mean results of the twenty
artificial datasets.

The results in Fig.~\ref{fig:artinocorr} show evidence for both
overestimation and underestimation of input values, dependent on the
parameter and cohort. Bias is most severe in the determination of
the change in power and supply rate in cohort\,\#\,1. In spite of
the offsets, it is encouraging to note that there is no appreciable
variation of the bias with degree. Any variation would certainly be
well within the uncertainties expected from analysis of a single,
3456-d dataset (which are a factor $\sqrt{20}$ larger than the
errors plotted in Fig.~\ref{fig:artinocorr}).

From these artificial data we then went on to derive a
window-function parameter correction that could be applied to the
real BiSON observations. Because of the small errors on the
artificial data results, the correction would be well constrained,
and uncertainties on the results in Fig.~\ref{fig:artinocorr} would
therefore not add significantly to the uncertainties on the final,
corrected BiSON results.

 \subsection{Formulation of correction procedure}
 \label{sec:corr}

The following procedure was applied to the fitted parameters within
each cohort of artificial data.

We began by computing the ratio of the fitted and input parameters
of every fitted mode. Ratios were made for the heights, $h$, and
widths, $\gamma$. Recall that each of the twenty independent 3456-d
sets was split into thirty-two 108-d pieces for fitting. We
therefore obtained many ratios, $Q_{h}(i,j,n,\ell)$ and
$Q_{\gamma}(i,j,n,\ell)$. The variable $i$ ran over the independent
sets ($1 \le i \le 20$); while $j$ ran over the 108-d pieces ($1 \le
j \le 32$), each having their own distinct BiSON window function.

Next, we averaged the ratios over the twenty independent datasets
(over $i$). This gave a total of thirty-two mean parameter ratios
for every mode, one for each 108-d window function. These mean
parameter ratios, $Q(j,n,\ell)$, were then regressed linearly
against the duty cycles, $D(i)$, of each of the windows; in fact, we
fitted to $1-D(i)$, rather than $D(i)$. The fits gave best-fitting
gradients $\delta Q^{\rm D}(n,\ell)$.

By following the above procedure for each cohort of artificial data,
we obtained two sets of mean ratios and best-fitting gradients. We
were then in a position to construct, from the ratios and gradients,
a formalism for correcting the fitted BiSON parameters for the
window function. In short, it would allow us to make estimates of
the unbiased parameters by linear extrapolation to $D=1.0$.

First, we checked the procedure could be applied to the fitted
artificial mode parameters to recover correct estimates of the mean
input changes in cohorts\,\#\,1 and\,\#\,2. This would involve
performing a full analysis on each of the artificial 3456-d sets,
and would be a dry run of the full correction and analysis as it
would be applied to real BiSON data. Taking the width parameter as
an example, estimates of the unbiased widths,
$\gamma_{Q}(j,n,\ell)$, were given by the linear combination:
 \begin{equation}
 \gamma_{Q}(j,n,\ell)=\gamma(j,n,\ell)\times\{Q_{\gamma}(j,n,\ell)+[1-D(i)]
 \times \delta Q^{\rm D}_\gamma(n,\ell)\}^{-1}.
 \label{eq:ycorr}
 \end{equation}
A similar equation was used to define a correction for the mode
heights. Once we had obtained estimates of the unbiased widths (and
heights) from Equation~\ref{eq:ycorr} (and its mode-height
counterpart), the corrected parameters were then used to estimate
mean parameter variations. We used the corrected parameters as input
to the analysis outlined in Section~\ref{sec:fitflux}, and found we
were able to recover, within errors, the input cycle shifts when we
analyzed datasets from cohort\,\#\,1; and the expected null changes
when we analyzed datasets from cohort\,\#\,2. To illustrate the
success of the correction procedure, Table~\ref{table:lcorr} shows
the linear correlation between the estimated and input width changes
for sets in cohort\,\#\,1 before (top row) and after (bottom row)
the correction was used. Given the numbers of data involved, these
represent significant improvements in the correlation, and the final
result.

\begin{table}[ht]
\centering \caption{Linear correlations between the fractional
variations in mode width (\% per unit of change in $F_{10.7}$)
extracted from the artificial datasets in cohort\#\,1 and the true,
input variations without and with temporal window function
corrections (Equation~\ref{eq:ycorr}).}
\begin{tabular}{c c c c c c c c c}\hline \hline
$\delta\gamma$  & $\ell=0$ & $\ell=1$ & $\ell=2$ & $\ell=3$ \\ \hline
No correction   & 0.80 & 0.93 & 0.96 & 0.61  \\
With correction & 0.98 & 0.99 & 0.99 & 0.94  \\ \hline
\end{tabular}
\label{table:lcorr}
\end{table}

 \section{Results with Sun-as-a-star BiSON data}
 \label{sec:bison}

With the correction checked on the artificial data, we proceeded to
apply the procedure to fitted parameters from the thirty-two 108-d
BiSON datasets. We used the mean parameter ratios, $Q(j,n,\ell)$,
and the best-fitting gradients, $\delta Q^{\rm D}(n,\ell)$, from the
artificial data. Recall there were two sets of these coefficients:
one from artificial cohort\,\#\,1, the other from artificial
cohort\,\#\,2. The window-function correction procedure was
therefore applied twice, yielding two sets of corrected parameters.

\begin{figure*}
\centering
\resizebox{\hsize}{!}{\includegraphics{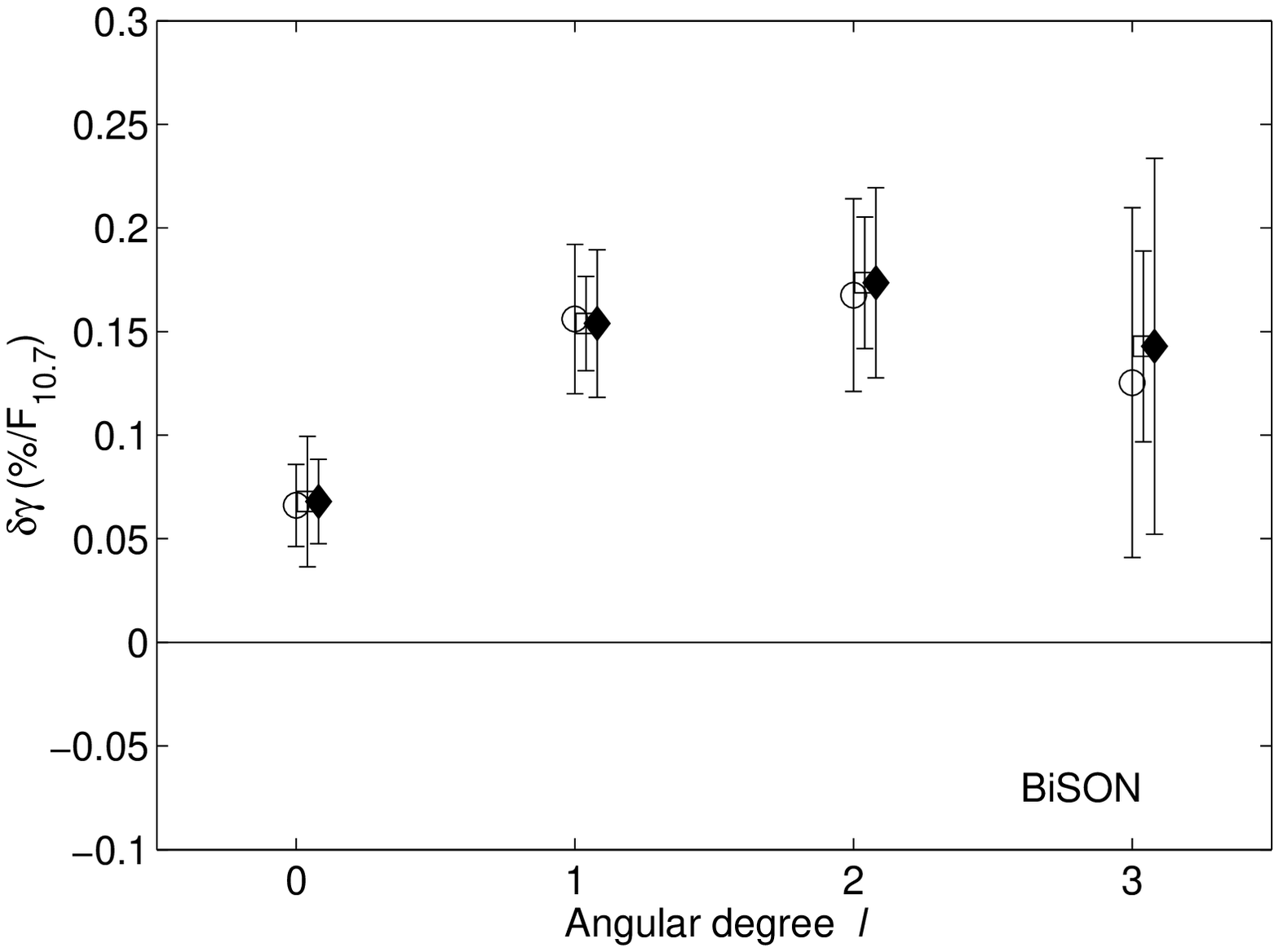}
                      \includegraphics{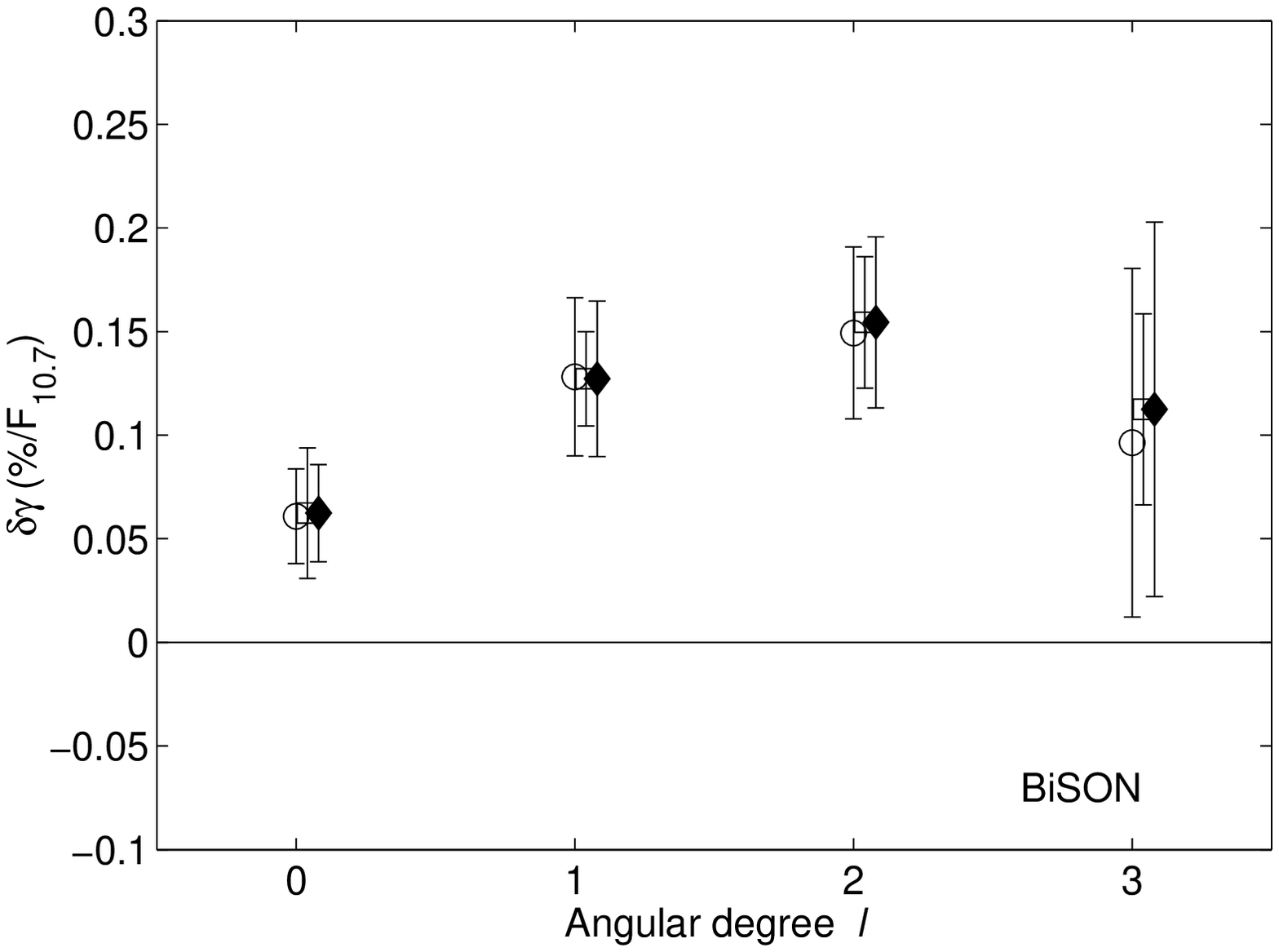}}
\resizebox{\hsize}{!}{\includegraphics{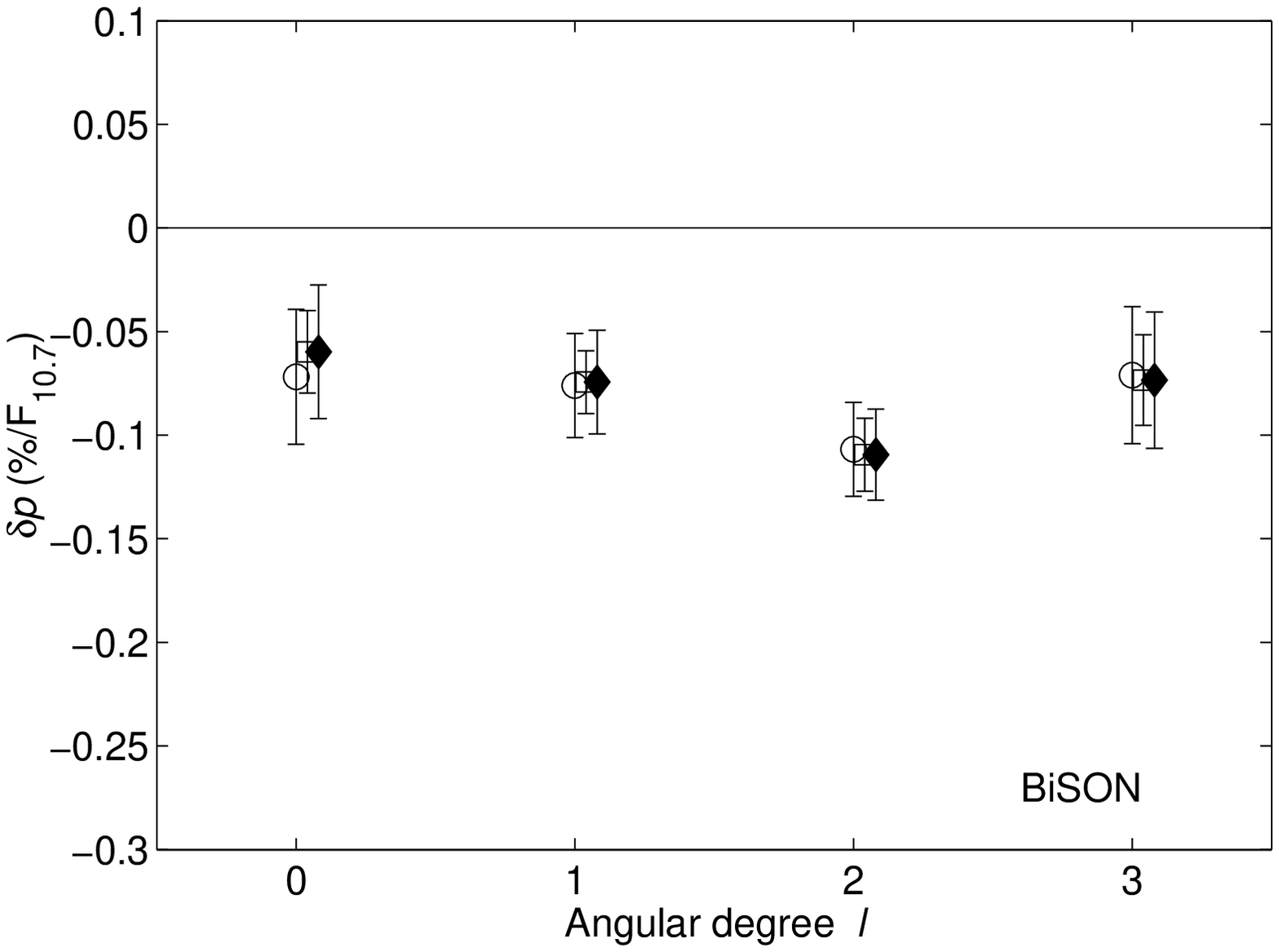}
                      \includegraphics{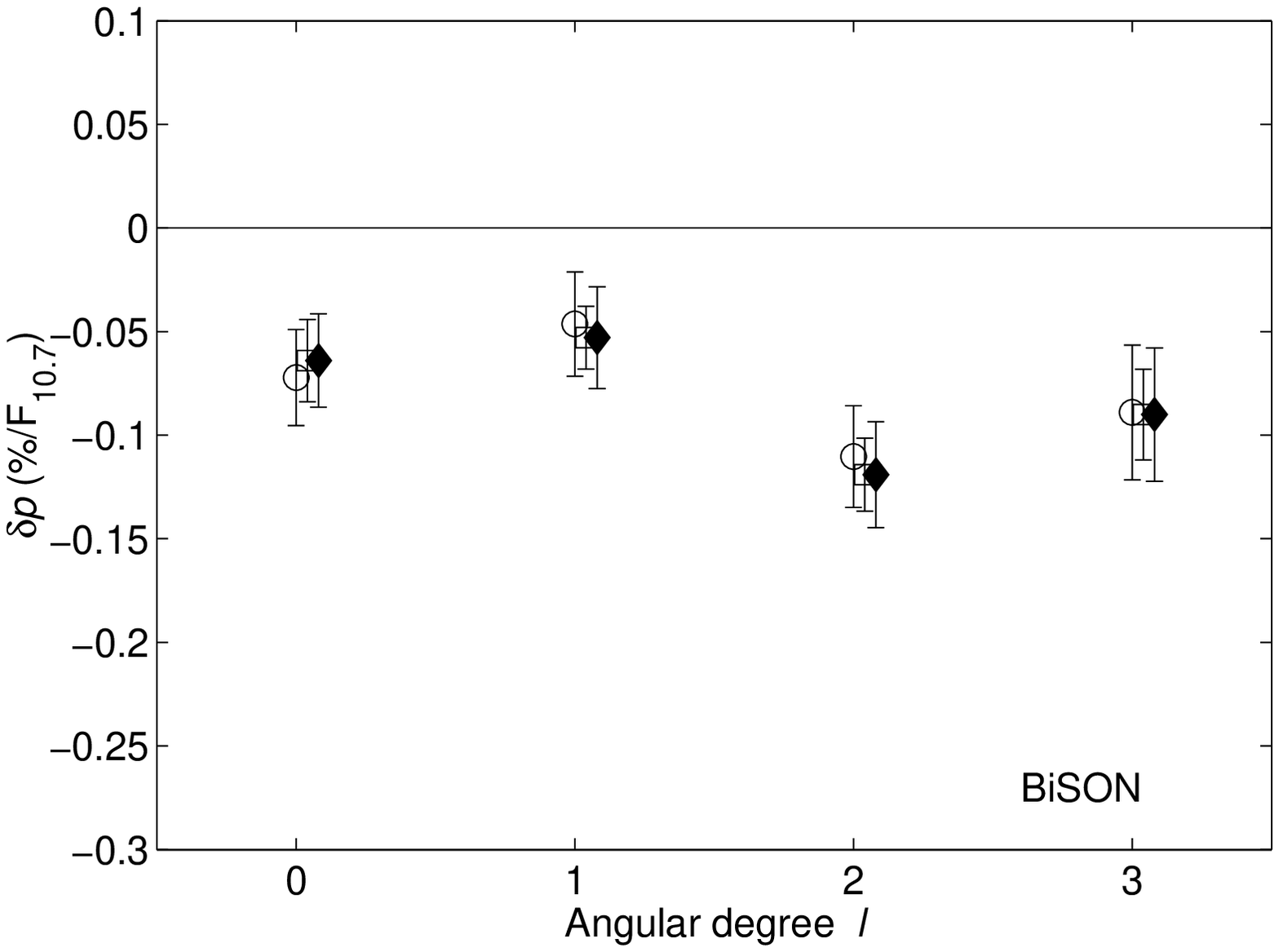}}
\resizebox{\hsize}{!}{\includegraphics{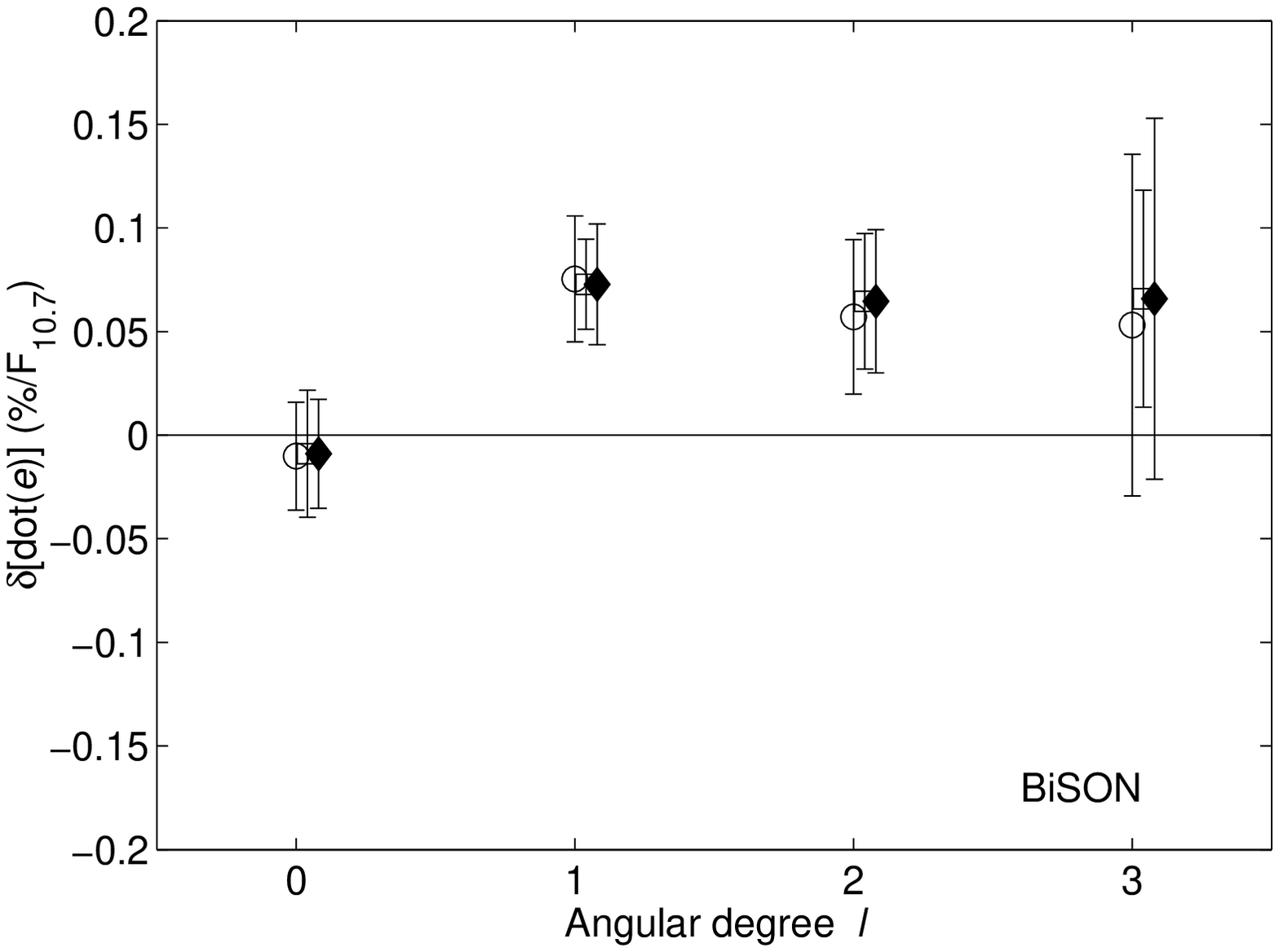}
                      \includegraphics{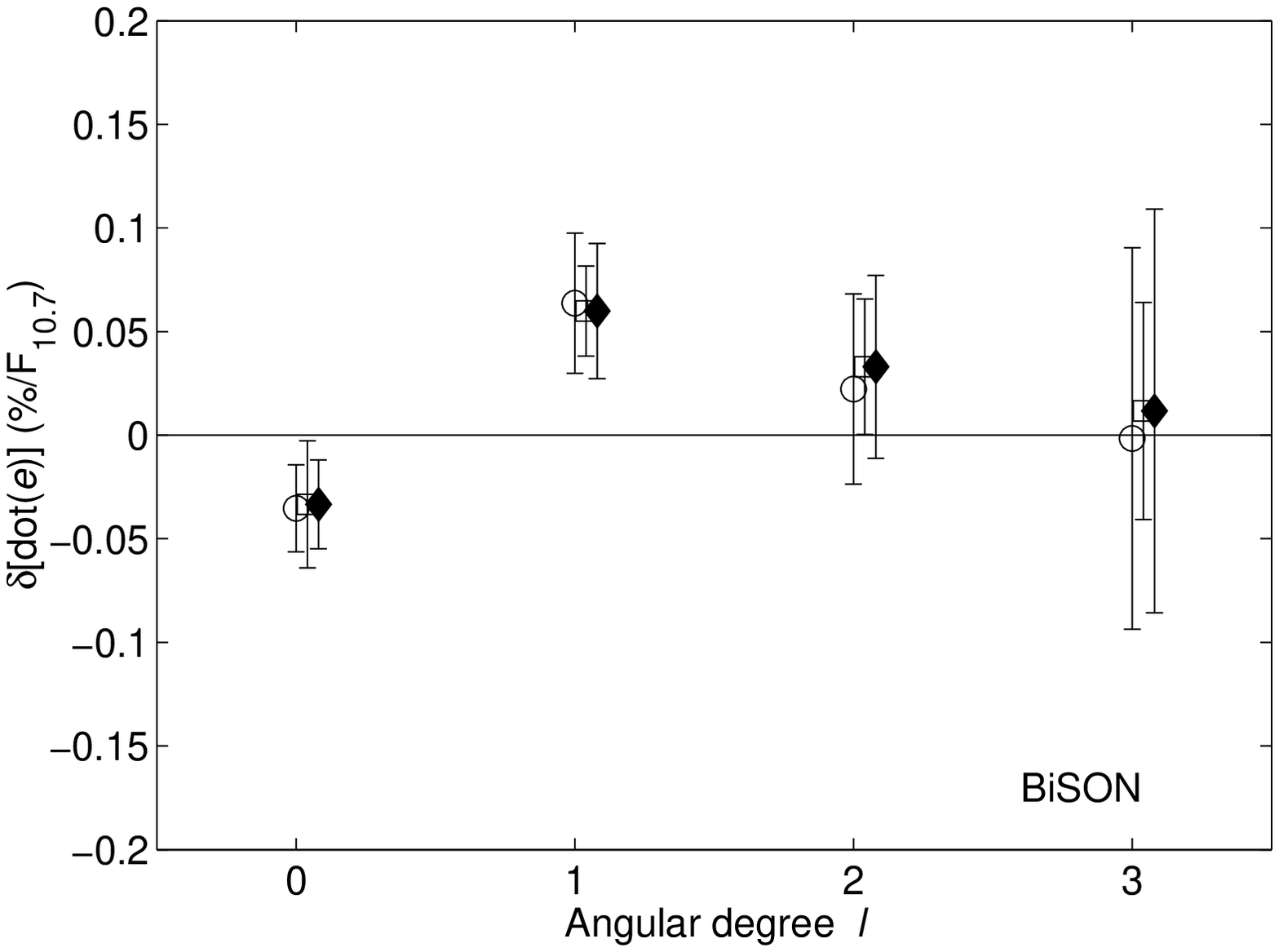}}

\caption{Results of analysis of BiSON data. Shown are mean
fractional variations (\% per unit of change in $F_{10.7}$) in mode
damping, $\delta \gamma$ ({\it upper panels}), mode velocity power,
$\delta p$ ({\it middle panels}), and mode energy supply rate,
$\delta \dot{e}$ ({\it lower panels}), as a function of the mode
angular degree $\ell$. Panels in the left-hand column show results
given using the cohort\,\#\,1 coefficients in the window-function
correction; those in the right-hand column results from using the
coefficients from cohort\,\#\,2. The solid horizontal lines in each
plot mark the zero-change level for reference. The symbols give
information on how the averages were computed (see
Sec.~\ref{sec:fitflux}): unweighted mean -$\circ$-; weighted mean
with internal error -$\square$-; weighted mean with external error
-$\blacklozenge$-.}
\label{fig:fracobs}
\end{figure*}

Estimates of the unbiased height and width parameters were then
analyzed, as per Section~\ref{sec:fitflux}. This give the sought-for
estimates of the mean parameter changes, per unit change in the
radio flux. Again, these changes -- which are shown in
Fig.~\ref{fig:fracobs} -- represent averages for each $\ell$ over
the range 2600 to $3600\,\rm \mu Hz$. The plots in the left-hand
column show results given by the cohort\,\#\,1 correction
coefficients; those in the right-hand panels results given by the
cohort\,\#\,2 coefficients. The results are also summarized in
Table~\ref{table:results}. Here, we have multiplied the
shift-per-unit-activity values shown in Fig.~\ref{fig:fracobs} (the
external mean data -$\blacklozenge$-) by the change in the radio
flux observed across the thirty-two 108-d pieces, to give estimates
of the mean parameter changes over the solar cycle.

We found good agreement, to within the errors, between results given
by the cohort\,\#\,1 and\,\#\,2 corrections. Averages made over the
$l=0$, 1 and 2 results were in reasonable agreement with previous
analyses of the BiSON, IRIS and GOLF Sun-as-a-star data
\citep{chaplin00,salabert03,chano03,chano04}. However, the most
striking feature of Fig.~\ref{fig:fracobs} is the apparent relative
weakness of changes in the $\ell=0$ widths, implying an $\ell$
dependence of the results. Inspection of results for the power
parameter showed hints of a mirrored, albeit weaker, trend. There
were also indications of a jump in $\ell$ in the supply-rate results
(in particular for the cohort\,\#\,1 corrected data).

To ascribe some formal significance to the weakness of the $\ell=0$
width shift, we first compared its value with the weighted mean of
the $\ell=1$ and 2 width shifts (we ignored the
less-well-constrained $\ell=3$ data). The $\ell=0$ value was found
to be about $2\sigma$ different from the mean over $\ell=1$ and 2,
for both cohort corrections. However, when we computed a linear
regression of the mean shifts on $\ell$, we found gradients $\approx
3\sigma$ and $\approx 4\sigma$ different from zero for the
cohort\,\#\,1 and cohort\,\#\,2 coefficients respectively. We
conclude that there is marginally significant evidence for an $\ell$
dependence of the width changes. Similar analysis of the other
parameters gave weaker levels of correlation, and therefore showed
no evidence for significant variation, with $\ell$. Whether the hint
of variation in the power and supply rate results is real only more
data, and independent confirmation from other observations (e.g.,
GOLF) will tell.

\begin{table*}[t]
\centering

\caption{Fractional variations (\%) of the low-degree p-mode damping
and excitation parameters from solar minimum to solar maximum for
each $\ell$ value, averaged over 2600 and 3600~$\mu$Hz. Final
results are shown for application of the window-function correction
coefficients from artificial data cohorts\,\#\,1 and \#\,2 (see
Sec.~\ref{sec:corr}). Data correspond to the external mean
shifts, rendered as -$\blacklozenge$-  symbols in
Fig.~\ref{fig:fracobs}.}

\begin{tabular}{cccccc}\hline \hline
 & & & & & \\
Parameter& $\ell=0$  & $\ell=1$  & $\ell=2$  & $\ell=3$  &
$\left<\ell=0,1,2\right>$ \\
 & & & & & \\
\hline
 & & & & & \\
\multicolumn{6}{c}{Correction coefficients from cohort\,\#\,1}\\
 & & & & & \\
$\delta\gamma$
           & $9.9\pm3.0\%$   & $22.5\pm5.2\%$  & $25.4\pm6.7\%$  & $20.9\pm13.3\%$ & $14.7\pm2.4\%$ \\
$\delta p$
           & $-8.7\pm4.7\%$  & $-10.9\pm3.7\%$  & $-16.0\pm3.2\%$ & $-10.7\pm4.8\%$ & $-12.7\pm2.2\%$ \\
$\delta \dot{e}$
           & $-1.3\pm3.8\%$   & $10.6\pm4.3\%$  & $9.5\pm5.1\%$  & $9.6\pm12.7\%$   & $ 5.2\pm2.5\%$\\
 & & & & & \\
\multicolumn{6}{c}{Correction coefficients from cohort\,\#\,2}\\
 & & & & & \\
$\delta\gamma$
           & $9.1\pm3.4\%$   & $18.6\pm5.5\%$  & $22.6\pm6.0\%$  & $16.4\pm13.2\%$ & $13.8\pm2.6\%$ \\
$\delta p$
           & $-9.4\pm3.3\%$  & $-7.7\pm3.6\%$  & $-17.4\pm3.7\%$ & $-13.2\pm4.7\%$ & $-11.3\pm2.0\%$ \\
$\delta \dot{e}$
           & $-4.9\pm3.1\%$   & $8.8\pm4.8\%$  & $4.8\pm6.4\%$  & $1.7\pm14.2\%$   & $ -0.1\pm2.4\%$\\
 & & & & & \\
           \hline

\end{tabular}
\label{table:results}
\end{table*}

 \section{Conclusion and discussion}

We have analyzed some 9.5-yr of BiSON Sun-as-a-star data --
collected in solar cycles 22 and 23 -- to search for dependence of
the mode excitation and damping parameter changes on the angular
degree, $\ell$, of the data. The nature of the Sun-as-a-star
observations is such that for changes measured at fixed frequency,
or for changes averaged across the same range in frequency, any
$\ell$ dependence present carries information on the latitudinal
distribution of the agent (i.e., the activity) responsible for those
changes.

We split the 9.5-yr timeseries into contiguous 108-d pieces, and
determined mean changes in the damping of, power in, and energy
supplied to, the modes through the solar cycle. We also applied a
careful correction to account for the deleterious effects of the
ground-based BiSON window function on the results. This correction
was calibrated by, and then fully tested on, artificial seismic data
generated by the solarFLAG mode simulation code.

From our full analysis we obtained a marginally significant result
for the damping parameter, where the mean change was found to be
weakest at $\ell=0$, and higher in data on the other $\ell$. The
result implies the damping is strongest in the active regions,
confirming results on the more numerous higher-$\ell$ data
\citep{komm02}. The other excitation and damping parameters we
investigated showed hints of some dependence in $\ell$, but nothing
that could judged as statistically significant.

Our main conclusion is that the mean fractional solar-cycle change
in the $\ell=0$ damping rates is approximately 50\,\% smaller than
was previously assumed. It had been common practice to use an
average over all low-$\ell$ modes (where mean solar-cycle shift
values have averaged about 18\,\%
\citep{chaplin00,salabert03,chano03,chano04}. Our downward revision
of the radial-mode value has implications for comparisons with
models of the global solar cycle changes, which are usually based on
a spherically symmetric geometry.

 \begin{acknowledgements}

This paper utilizes data collected by the Birmingham
Solar-Oscillations Network (BiSON), which is funded by the UK
Particle Physics and Astronomy Research Council (PPARC). We thank
the members of the BiSON team, colleagues at our host institutes,
and all others, past and present, who have been associated with
BiSON. The GOLF instrument is the result of cooperative endeavour of
many individuals, to whom we are deeply indebted. SOHO is a mission
of international cooperation between \textit{ESA} and \textit{NASA}.
The 10.7-cm radio flux observations are made at Penticton by the
National Research Council of Canada and are available from the World
Data Center.

 \end{acknowledgements}

\end{document}